\documentclass[preprint,showpacs,preprintnumbers,amsmath,amssymb,aps]{revtex4}

\usepackage{graphicx}
\usepackage{bm}

\begin{document}


\title{Improved transfer efficiency with pulsed atom transfer between two magneto-optical traps}
\author{S P Ram}
\email{spram@rrcat.gov.in}
\author{S K Tiwari} 
\author{S R Mishra}
\affiliation{%
Laser Physics Applications Division, Raja Ramanna Centre for Advanced Technology, Indore 452013 India.}%


\begin{abstract}
In our double magneto-optical trap (MOT) setup containing a vapor chamber MOT (VC-MOT) and an ultra high vacuum MOT (UHV-MOT) for $^{87}$Rb atoms, we find that transfer of atoms from VC-MOT to UHV-MOT can be enhanced by employing a pulsed VC-MOT loading followed by a pulsed push beam, as compared to that obtained by focusing a continuous wave (CW) push beam on a continuously loaded VC-MOT. By choosing appropriately the VC-MOT duration and push beam duration, the number of atoms in UHV-MOT was $\sim$3-times the number obtained with a continuous VC-MOT and a CW push beam of optimized power. The processes affecting the pulsed transfer have been studied.
\end{abstract}

\maketitle
%
To achieve Bose-Einstein condensation (BEC) in alkali atomic gases, laser cooled atoms in a magneto-optical trap (MOT) are trapped in a purely magnetic trap (or dipole trap) to perform evaporative cooling to lower temperature to reach critical value \cite{fermischool}. Since evaporative cooling is lossy and slow process, one needs a large number of atoms in magnetic trap with a long life-time of the trap. Since magnetic trap life-time is adversely affected by increase in vapor pressure in the chamber, required vapor pressure for a significant number in a MOT is detrimental to the formation of magnetic trap and evaporative cooling in the vapor MOT chamber. One option is to use double-MOT setup ~\cite{myatt, gibble}, in which two magneto-optical traps are formed at different vacuum level. First MOT is formed in a chamber containing atomic vapor (called VC-MOT) at pressure of $\sim10^{-8}$ Torr, and second MOT (i.e. UHV-MOT) is formed in a UHV chamber (at pressure of $\sim10^{-10}$ to $10^{-11}$ Torr). The UHV-MOT provides atoms for magnetic trap and evaporative cooling in which atoms are accumulated after transfer from VC-MOT cloud. Various methods such as moving magnetic coils ~\cite{cornell,jin}, imbalance of trapping beams forces ~\cite{park}, or use of a push beam ~\cite{myatt, swanson, cacciapuoti} are employed to transfer atoms from VC-MOT to UHV-MOT. Among these, use of a push beam is simple and efficient.   

Efficient transfer of atoms from VC-MOT to UHV-MOT is clearly of prime importance in the double-MOT systems. Use of various auxilary techniques (such as use of guiding magnetic field ~\cite{myatt}, two-dimensional cooling ~\cite{swanson}, a guiding laser beam ~\cite{dimova} and a hollow laser beam ~\cite{mishra2} during the atom transfer) along with a push beam have been reported to enhance transfer efficiency. Among early experiments on transfer, \citet{cacciapuoti} used a focused continuous wave (CW) push beam to create an extraction column in VC-MOT to obtain a continuous transfer between two magneto-optical traps (MOTs). In their experiments, they found that use of CW push beam was more efficient than the pulsed push beam. In the present work, we find that repetitive use of a pulsed (and unfocused) push beam interacting with a partially loaded VC-MOT cloud results in more number of atoms transferred to UHV-MOT, as compared to that obtained with a CW push beam employed on a continuous VC-MOT. The processes affecting the transfer in the above pulsed scheme have been investigated. These results can guide one to optimize various parameters, such as VC-MOT duration and push beam duration, for efficient transfer of atoms from VC-MOT to UHV-MOT to accumulate large number of atoms in UHV-MOT. 

\begin{figure}
	\centering
		\includegraphics[scale = 0.5]{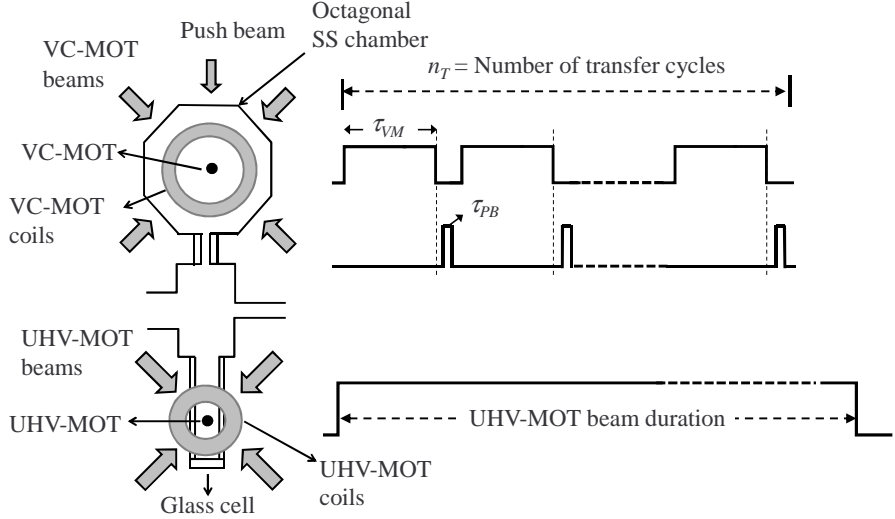}
	\caption{Schematic of the experimental setup and sequence of pulses for pulsed push beam transfer scheme. ${\tau}_{_{VM}}$: duration of VC-MOT and ${\tau}_{_{PB}}$: push beam duration.}
	\label{fig:fig1}
\end{figure}
Fig. \ref{fig:fig1} shows schematic of our experimental setup which consists of vacuum system in two parts. First part is an octagonal chamber of stainless-steel at a pressure of $\sim2\times 10^{-8}$ Torr (with Rb-vapor) in which VC-MOT is formed. The second part is a glass cell maintained at a pressure of $\sim2\times10^{-10}$ Torr in which UHV-MOT is formed. These two parts of the system are connected by a narrow tube (8 mm diameter and 70 mm length) which maintains the pressure difference between two parts due to differential pumping and allows transfer of atoms between MOTs formed in these two parts. The Rb-vapor for VC-MOT in octagonal chamber was generated by passing a DC current of 3.5 A in a Rb-getter source. The cooling laser beams for both the MOTs were generated from two separate grating controlled external cavity diode lasers, whereas the re-pumping laser beams were obtained by dividing output from a single laser of similar configuration. For VC-MOT, the cooling beam was first passed through two acousto-optic modulators (AOMs) controlling its duration, and then expanded and split into three beams (each having $\sim$7 mW power and 1/e$^2$ radius of $\sim$4 mm ) which entered the chamber. For splitting of beam, we used a combination of half-wave plates and polarizing beam splitter (PBS) cubes. After retro-reflection of each of these three beams, the desired six VC-MOT beams were obtained. The re-pumping laser beam for VC-MOT ($\sim$9 mW power ) was mixed with one of the three cooling beams entering the chamber. For UHV-MOT, the cooling and re-pumping beams were first mixed and then the combined beam was passed through two AOMs. After passing through AOMs, the beam was expanded (to 1/e$^2$ radius of $\sim$5 mm) and then divided into five UHV-MOT beams using a sequence of half-wave plate and PBS elements. It was ascertained that all UHV-MOT beams had nearly same power ($\sim$6 mW in each) in cooling part. However, power in re-pumping part in these beams was not equal due to different polarization of re-pumping laser beam. The total re-pumping power was $\sim$12 mW in all five beams. Four of these UHV-MOT beams were used as two pairs of counter-propagating beams incident at 45 degree on the glass cell surface. This was done because retro-reflection arrangement at this incidence angle would result in large power difference between forward propagating and retro-reflected beams. The fifth beam, which was at normal incidence to glass cell surface, was used in retro-reflection arrangement to have fifth and sixth beams for UHV-MOT. 

All lasers were locked using well known saturated absorption spectroscopy (SAS). The cooling laser was locked at side ($\sim$12 MHz to the red) of 5S$_{1/2}$ F = 2$\rightarrow$ 5P$_{3/2}$F$^\prime$ = 3 transition of $^{87}$Rb, whereas re-pumping laser was locked at peak of 5S$_{1/2}$ F = 1$\rightarrow$ 5P$_{3/2}$F$^\prime$ = 2 transition of $^{87}$Rb. Appropriate polarization of beams were set using quarter-wave retardation plates. Axial magnetic field gradients of $\sim$12 G/cm and $\sim$10 G/cm,for VC-MOT and UHV-MOT respectively, were generated by two pairs of quadrupole coils. The push beam (maximum power $\sim$4 mW) was obtained from a part of the VC-MOT cooling laser beam after shifting its frequency by two AOMs to peak of 5S$_{1/2}$ F = 2$\rightarrow$ 5P$_{3/2}$F$^\prime$ = 3 transition of $^{87}$Rb. The push beam was aligned vertically which propagated from VC-MOT chamber to UHV-MOT cell through differential pumping tube. Applying appropriate trigger to these AOMs, the push beam could be made either pulsed or CW, as per requirement. For pulsed atom transfer, the VC-MOT was loaded for a variable time duration (${\tau}_{_{VM}}$). Subsequent to VC-MOT loading, cooling beams were switched-off and pulse of push beam of duration ($\tau_{_{PB}}$) was applied to impart an impulse on VC-MOT atoms to transfer them to UHV-MOT. This was repeated for a number of transfer cycles ($n_T$) (as shown in Fig. \ref{fig:fig1}) to accumulate atoms in the UHV-MOT (vertically below) for which cooling and trapping MOT-beams were continuously on. We estimated the number of atoms in the MOTs by the fluorescence imaging technique described earlier \cite{mishra2}.
\begin{figure}
	\centering
		\includegraphics[scale = 1.0]{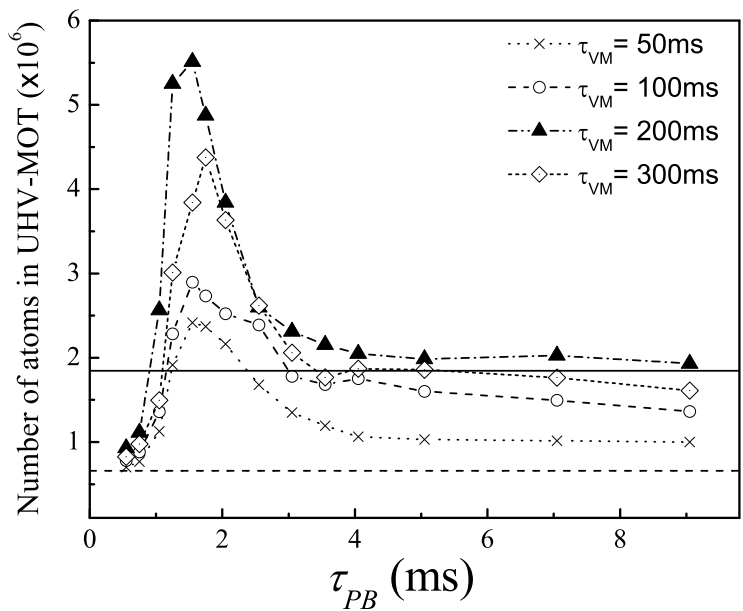}
	\caption{ Variation in number of atoms in UHV-MOT with $\tau_{_{PB}}$ (duration of unfocused push beam) for different VC-MOT loading duration $\tau_{_{VM}}$. Continuous horizontal line and dashed line shows the number of atoms in UHV-MOT for a focused CW push beam and unfocused CW push beam, respectively, of optimized powers.}
	\label{fig:fig2}
\end{figure}

In our VC-MOT, maximum number of atoms that could be accumulated was $\sim7\times10^7$ and this took $\sim$1200 ms of VC-MOT loading time. We aligned push beam in vertical direction such that it was interacting with VC-MOT cloud and just missing the UHV-MOT cloud. In the absence of push beam the cloud was formed only in VC-MOT, and atom cloud in UHV-MOT was formed when push beam was applied on VC-MOT cloud to push atoms in downward direction. When we used an unfocused (1/e$^2$ radius of $\sim$ 2.0 mm) CW push beam, atom cloud in both the MOTs disappeared at few mW power. The cloud in both the MOTs reappeared when push beam power was reduced. For this push beam, the number in UHV-MOT was maximum at an optimum power of $\sim$500 $\mu$W. At higher power, unfocused push beam which covers a large VC-MOT volume prevents the formation of cloud in VC-MOT due to imbalance of the trapping forces. When we focused the CW push beam on VC-MOT, its effect was different than that of unfocused CW push beam. A focused push beam imbalances trapping forces in the smaller region of VC-MOT volume, and allows a better loading of VC-MOT with a small leak in it giving a flux of atoms from VC-MOT to UHV-MOT. With focused (1/e$^2$ radius $\sim$ 35 $\mu$m at VC-MOT cloud) CW push beam, the number in UHV-MOT could be reached higher than that could be reached with unfocused CW push beam. The maximum number ($\sim1.8\times10^6$) in UHV-MOT with focused push beam was obtained at push beam power of $\sim150\ \mu$W (shown in Fig. \ref{fig:fig2} by continuous horizontal line).

Next, we employed the above focused push beam in the pulsed form of variable duration with pulse sequences as shown in Fig. \ref{fig:fig1}. The number of atoms in UHV-MOT with this push beam was found to be smaller than the maximum number obtained with focused CW push beam, which was tried for different combinations of push beam intensity and pulse duration and for different number of transfer cycles ($n_T$) upto $n_T$ =100. This observation is similar to the result of \citet{cacciapuoti}, who observed a better loading of UHV-MOT with a focused CW push beam than that with a focused pulsed push beam. With a pulsed push beam, it is desirable to switch-off VC-MOT beams (to switch-off trapping potential) so that trapped atoms could easily (at less intensity) escape the MOT volume. A focused push beam (due to its higher divergence) gives more transverse acceleration to VC-MOT atoms than an unfocused beam. Due to absence of cooling MOT beams, this acceleration leads to a higher divergence of atom flux from VC-MOT which results in lower number in UHV-MOT. Thus focused push beam performs poorer in the pulse mode than in CW mode.
  
The above observations and understanding prompted us to use the unfocused push beam, instead of focused push beam, in the above pulsed mode of transfer. The push beam we used in these experiments had maximum power of $\sim$4 mW (i.e. maximum intensity of $\sim$60 mW/cm$^2$). With the use of unfocused push beam in pulsed mode, the number of atoms in UHV-MOT was increased as compared to focused push beam. The results of these measurements with unfocused pulsed push beam are shown in Fig. \ref{fig:fig2}, which shows number of atoms accumulated in UHV-MOT as function of push beam pulse duration ($\tau_{_{PB}}$) and for different values of VC-MOT loading duration($\tau_{_{VM}}$). For these measurements, we kept $n_T$=100 (number of transfer cycles) for each combination of $\tau_{_{PB}}$ and $\tau_{_{VM}}$ values in data in Fig. \ref{fig:fig2}. As is evident from these results, the number in UHV-MOT is dependent on push beam duration ($\tau_{_{PB}}$) and VC-MOT duration ($\tau_{_{VM}}$). The optimum values can be chosen to get the maximum number in UHV-MOT. As shown in the figure, the number of atoms in UHV-MOT was enhanced upto $\sim$3-times the number we obtain with the focused CW push beam of optimized power (horizontal line in Fig. \ref{fig:fig2}). 

This enhancement in UHV-MOT number with unfocused pulsed push beam (Fig. \ref{fig:fig2}) can be understood as result of following factors. First, a smaller divergence of unfocused beam results in a more collimated atom flux which results in a higher number of atoms reaching the UHV-MOT volume. Second, the unfocused push beam interacts with nearly entire VC-MOT cloud and can push larger number of atoms to UHV-MOT than the focused push beam. And finally, in pulsed mode transfer, VC-MOT loading is unhindered by push beam presence.

The observed optimum push beam duration ($\tau_{_{PB}} \sim$ 1.75 ms for peak intensity $\sim$60 mW/cm$^2$) in Fig. \ref{fig:fig2} can be correlated to velocity of transfer acquired by atoms due to push beam impulse and capture velocity of UHV-MOT. The optimum $\tau_{_{PB}}$ (or impulse) should result in transfer velocity of atoms comparable to the capture velocity of UHV-MOT \cite{muniz}. For lower transfer velocity, less number reaches to capture volume of UHV-MOT due to loss of atoms in transverse direction during transit. For transfer velocity higher than the capture velocity of UHV-MOT, atoms escape the UHV-MOT volume without being captured there.   

Another important parameter, in experimental results shown in Fig. \ref{fig:fig2}, is duration of VC-MOT ($\tau_{_{VM}}$).The duration of VC-MOT affects the number accumulated in UHV-MOT in the pulsed transfer scheme. In our experiments, we obtained the largest number in UHV-MOT for  $\tau_{_{VM}}$ = 200 ms as shown in Fig. \ref{fig:fig2}. We have investigated this dependence of number in UHV-MOT on $\tau_{_{VM}}$ in more detail. For a smaller value of $\tau_{_{VM}}$ ($<$ 200 ms), a relatively less number obtained in UHV-MOT seems a result of poor loading of VC-MOT giving less number of atoms available for transfer to UHV-MOT. But for larger values of $\tau_{_{VM}}$ ($>$200 ms), for which larger number of atoms (than for $\tau_{_{VM}}$ = 200 ms) are trapped in VC-MOT, the observed smaller number in UHV-MOT drew our attention to some other processes which might be playing the role. 

A larger $\tau_{_{VM}}$ implies a larger number in VC-MOT and a stronger fluorescence from VC-MOT. The fluorescence from VC-MOT can also be destructive to UHV-MOT, and hence can result in a smaller number in UHV-MOT for longer VC-MOT duration. Another factor is the dependency of temperature of VC-MOT cloud on $\tau_{_{VM}}$ (before the saturation) \cite{cooper}. A higher temperature of VC-MOT atoms would result in a larger transverse divergence of atom flux, and a smaller number of atoms captured in the UHV-MOT. We have investigated both these processes and found that they may be contributing to the observed reduction in number in UHV-MOT for larger $\tau_{_{VM}}$ ($>$ 200 ms).  

\begin{figure}
	\centering
		\includegraphics[scale = 1.0]{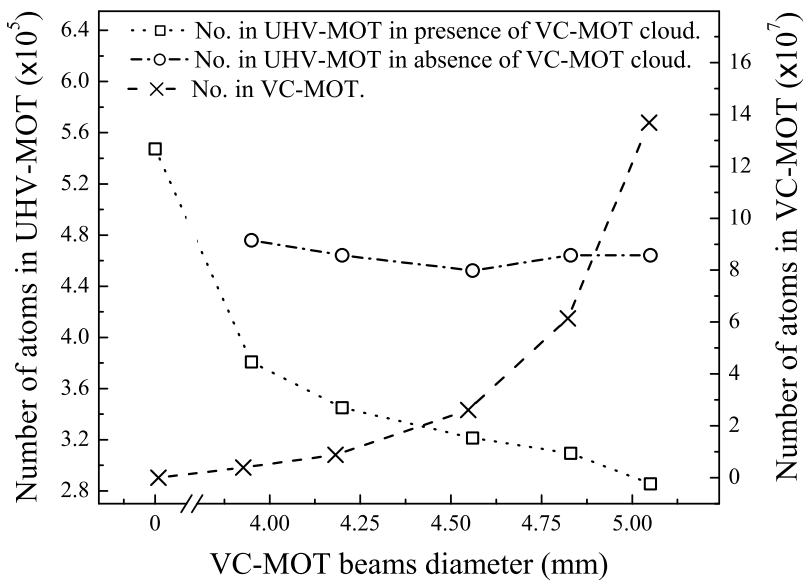}
	\caption{Variation in number of atoms in UHV-MOT and VC-MOT with diameter of VC-MOT beams (in absence of push beam and at a getter current 3.65A).}
	\label{fig:fig3}
\end{figure}

To study the destructive effect of VC-MOT fluorescence on UHV-MOT atoms, we studied the variation in number of atoms in UHV-MOT as function of number in VC-MOT, when both the MOTs were loaded independently from background vapor atoms. For this push beam was kept off, and getter-current was increased to 3.65 A so that background vapor level in UHV-MOT chamber became sufficient to load UHV-MOT. To vary number in VC-MOT, we varied only the diameter of VC-MOT beams while keeping other parameters fixed. For each diameter value of VC-MOT beams, we measured the number in VC-MOT as well as number in UHV-MOT. As shown in Fig. \ref{fig:fig3}, the number of atoms in UHV-MOT decreased with increase in VC-MOT beams diameter, whereas number in VC-MOT increased with diameter. To ensure that above variation in UHV-MOT number was indeed from variation in VC-MOT number and not from VC-MOT beams size, we also recorded the variation number in UHV-MOT (shown by circles in Fig. \ref{fig:fig3}) with VC-MOT beams diameter in absence of VC-MOT cloud (by switching-off current in VC-MOT coils). As is evident from Fig. \ref{fig:fig3}, this variation was insignificant. Thus number in UHV-MOT gets adversely affected by number in VC-MOT. This evidently seems to be due to resonant fluorescence from the VC-MOT cloud which is proportional to number in VC-MOT.
\begin{figure}
	\centering
		\includegraphics[scale = 1.0]{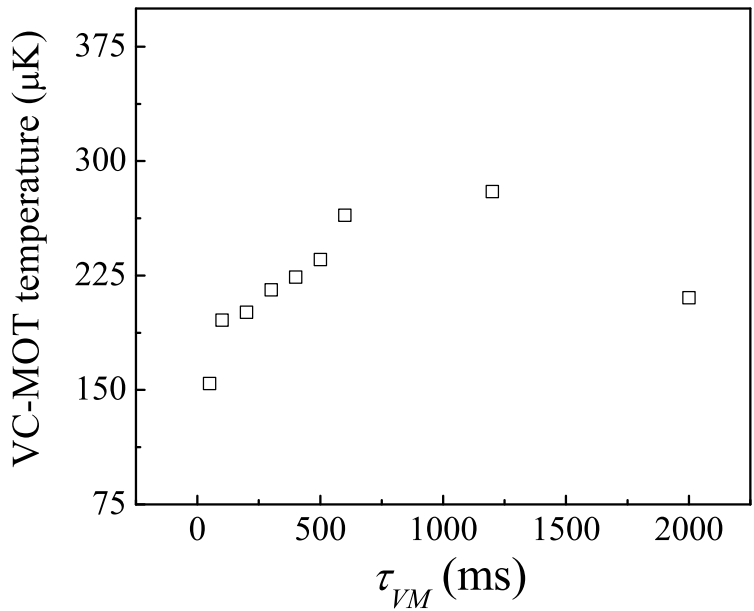}
	\caption{Variation of temperature of VC-MOT cloud with VC-MOT loading duration ($\tau_{_{VM}})$.}
	\label{fig:fig4}
\end{figure}

A larger $\tau_{_{VM}}$ also implies a larger number lost from UHV-MOT during two consecutive loadings of UHV-MOT, due to UHV-MOT life-time ($\sim$ few seconds). This is additive to loss due to VC-MOT fluorescence, during each transfer cycle. Both these effects combined may result the final number (accumulated after $n_T$ transfer cycles) in UHV-MOT smaller for a larger $\tau_{_{VM}}$.

We also measured the temperature of VC-MOT atom cloud as function of its loading duration $\tau_{_{VM}}$ (Fig. \ref{fig:fig4}). We note that for smaller ($<$1000 ms) values of $\tau_{_{VM}}$, temperature of cloud increased with $\tau_{_{VM}}$. A similar result has been reported earlier by \citet{cooper}. Thus, for VC-MOT duration ($\tau_{_{VM}}$) falling in this regime, the divergence of atom flux (and hence transfer efficiency) in the pulsed transfer experiments presented here is expected to be more for larger $\tau_{_{VM}}$. This higher divergence of atom flux is expected to result in poor loading of UHV-MOT with smaller number of atoms in it. 

We found that use of an unfocused pulsed push beam can be better than using a focused CW push beam, for efficient transfer of atoms from a VC-MOT to an UHV-MOT in a double-MOT setup. We investigated the dependence of number of atoms collected in UHV-MOT on push beam and VC-MOT parameters. Under optimum conditions, the observed number in UHV-MOT was $\sim$3-times larger than that obtained with a focused CW push beam of optimized power. These results would enable us to improve the number in UHV-MOT further for purpose of evaporative cooling to achieve BEC.


We thank S. C. Mehendale for a critical reading of the manuscript and helpful suggestions. We also thank  L. Jain, V. P. Bhanage, P. P. Deshpande, M. A. Ansari, H. R. Bundel and C. P. Navathe for development of the controller system for the setup and H. S. Vora for providing the image-processing software.


\begin{thebibliography}{12}
\expandafter\ifx\csname natexlab\endcsname\relax\def\natexlab#1{#1}\fi
\expandafter\ifx\csname bibnamefont\endcsname\relax
  \def\bibnamefont#1{#1}\fi
\expandafter\ifx\csname bibfnamefont\endcsname\relax
  \def\bibfnamefont#1{#1}\fi
\expandafter\ifx\csname citenamefont\endcsname\relax
  \def\citenamefont#1{#1}\fi
\expandafter\ifx\csname url\endcsname\relax
  \def\url#1{\texttt{#1}}\fi
\expandafter\ifx\csname urlprefix\endcsname\relax\def\urlprefix{URL }\fi
\providecommand{\bibinfo}[2]{#2}
\providecommand{\eprint}[2][]{\url{#2}}

\bibitem[{fer(1999)}]{fermischool}
\emph{\bibinfo{booktitle}{Proceedings of the International School of
  Physics, Enrico Fermi Course CXL}}, edited by
  \bibinfo{editor}{\bibfnamefont{M.}~\bibnamefont{Inguscio}},
  \bibinfo{editor}{\bibfnamefont{S.}~\bibnamefont{Stringari}},
  \bibnamefont{and} \bibinfo{editor}{\bibfnamefont{C.~E.} \bibnamefont{Wieman}}
  (\bibinfo{publisher}{IOS Press, Amsterdam}, \bibinfo{year}{1999}).

\bibitem[{\citenamefont{Myatt et~al.}(1996)\citenamefont{Myatt, Newbury,
  Ghrist, Loutzenhiser, and Wieman}}]{myatt}
\bibinfo{author}{\bibfnamefont{C.~J.} \bibnamefont{Myatt}},
  \bibinfo{author}{\bibfnamefont{N.~R.} \bibnamefont{Newbury}},
  \bibinfo{author}{\bibfnamefont{R.~W.} \bibnamefont{Ghrist}},
  \bibinfo{author}{\bibfnamefont{S.}~\bibnamefont{Loutzenhiser}},
  \bibnamefont{and} \bibinfo{author}{\bibfnamefont{C.~E.}
  \bibnamefont{Wieman}}, \bibinfo{journal}{Opt. Lett.}
  \textbf{\bibinfo{volume}{21}}, \bibinfo{pages}{290} (\bibinfo{year}{1996}).

\bibitem[{\citenamefont{Gibble et~al.}(1995)\citenamefont{Gibble, Chang, and
  Legere}}]{gibble}
\bibinfo{author}{\bibfnamefont{K.}~\bibnamefont{Gibble}},
  \bibinfo{author}{\bibfnamefont{S.}~\bibnamefont{Chang}}, \bibnamefont{and}
  \bibinfo{author}{\bibfnamefont{R.}~\bibnamefont{Legere}},
  \bibinfo{journal}{Phys. Rev. Lett.} \textbf{\bibinfo{volume}{75}},
  \bibinfo{pages}{2666} (\bibinfo{year}{1995}).

\bibitem[{\citenamefont{Lewandowski et~al.}(2003)\citenamefont{Lewandowski,
  Harber, Whitaker, and Cornell}}]{cornell}
\bibinfo{author}{\bibfnamefont{H.~J.} \bibnamefont{Lewandowski}},
  \bibinfo{author}{\bibfnamefont{D.~M.} \bibnamefont{Harber}},
  \bibinfo{author}{\bibfnamefont{D.~L.} \bibnamefont{Whitaker}},
  \bibnamefont{and} \bibinfo{author}{\bibfnamefont{E.~A.}
  \bibnamefont{Cornell}}, \bibinfo{journal}{J. Low. Temp. Phys.}
  \textbf{\bibinfo{volume}{132}}, \bibinfo{pages}{309} (\bibinfo{year}{2003}).

\bibitem[{\citenamefont{Goldwin et~al.}(2004)\citenamefont{Goldwin, Inouye,
  Olsen, Newman, DePaola, and Jin}}]{jin}
\bibinfo{author}{\bibfnamefont{J.}~\bibnamefont{Goldwin}},
  \bibinfo{author}{\bibfnamefont{S.}~\bibnamefont{Inouye}},
  \bibinfo{author}{\bibfnamefont{M.~L.} \bibnamefont{Olsen}},
  \bibinfo{author}{\bibfnamefont{B.}~\bibnamefont{Newman}},
  \bibinfo{author}{\bibfnamefont{B.~D.} \bibnamefont{DePaola}},
  \bibnamefont{and} \bibinfo{author}{\bibfnamefont{D.~S.} \bibnamefont{Jin}},
  \bibinfo{journal}{Phys. Rev. A} \textbf{\bibinfo{volume}{70}},
  \bibinfo{pages}{021601} (\bibinfo{year}{2004}).

\bibitem[{\citenamefont{Park et~al.}(1999)\citenamefont{Park, Jun, and
  Cho}}]{park}
\bibinfo{author}{\bibfnamefont{C.~Y.} \bibnamefont{Park}},
  \bibinfo{author}{\bibfnamefont{M.~S.} \bibnamefont{Jun}}, \bibnamefont{and}
  \bibinfo{author}{\bibfnamefont{D.}~\bibnamefont{Cho}}, \bibinfo{journal}{J.
  Opt. Soc. Am. B} \textbf{\bibinfo{volume}{16}}, \bibinfo{pages}{994}
  (\bibinfo{year}{1999}).

\bibitem[{\citenamefont{Swanson et~al.}(1998)\citenamefont{Swanson, Asgeirsson,
  Behr, Gorelov, and Melconian}}]{swanson}
\bibinfo{author}{\bibfnamefont{T.~B.} \bibnamefont{Swanson}},
  \bibinfo{author}{\bibfnamefont{D.}~\bibnamefont{Asgeirsson}},
  \bibinfo{author}{\bibfnamefont{J.~A.} \bibnamefont{Behr}},
  \bibinfo{author}{\bibfnamefont{A.}~\bibnamefont{Gorelov}}, \bibnamefont{and}
  \bibinfo{author}{\bibfnamefont{D.}~\bibnamefont{Melconian}},
  \bibinfo{journal}{J. Opt. Soc. Am. B} \textbf{\bibinfo{volume}{15}},
  \bibinfo{pages}{2641} (\bibinfo{year}{1998}).

\bibitem[{\citenamefont{Cacciapuoti et~al.}(2001)\citenamefont{Cacciapuoti,
  Castrillo, de~Angelis, and Tino}}]{cacciapuoti}
\bibinfo{author}{\bibfnamefont{L.}~\bibnamefont{Cacciapuoti}},
  \bibinfo{author}{\bibfnamefont{A.}~\bibnamefont{Castrillo}},
  \bibinfo{author}{\bibfnamefont{M.}~\bibnamefont{de~Angelis}},
  \bibnamefont{and} \bibinfo{author}{\bibfnamefont{G.}~\bibnamefont{Tino}},
  \bibinfo{journal}{Eur. Phys. J. D} \textbf{\bibinfo{volume}{15}},
  \bibinfo{pages}{245} (\bibinfo{year}{2001}).

\bibitem[{\citenamefont{Dimova et~al.}(2007)\citenamefont{Dimova, Morizot,
  Stern, Alzar, Fioretti, Lorent, Comparat, Perrin, and Pillet}}]{dimova}
\bibinfo{author}{\bibfnamefont{E.}~\bibnamefont{Dimova}},
  \bibinfo{author}{\bibfnamefont{O.}~\bibnamefont{Morizot}},
  \bibinfo{author}{\bibfnamefont{G.}~\bibnamefont{Stern}},
  \bibinfo{author}{\bibfnamefont{C.~G.} \bibnamefont{Alzar}},
  \bibinfo{author}{\bibfnamefont{A.}~\bibnamefont{Fioretti}},
  \bibinfo{author}{\bibfnamefont{V.}~\bibnamefont{Lorent}},
  \bibinfo{author}{\bibfnamefont{D.}~\bibnamefont{Comparat}},
  \bibinfo{author}{\bibfnamefont{H.}~\bibnamefont{Perrin}}, \bibnamefont{and}
  \bibinfo{author}{\bibfnamefont{P.}~\bibnamefont{Pillet}},
  \bibinfo{journal}{Eur. Phys. J. D} \textbf{\bibinfo{volume}{42}},
  \bibinfo{pages}{299} (\bibinfo{year}{2007}),
  \bibinfo{note}{arXiv:quant-ph/0609208v2}.

\bibitem[{\citenamefont{Mishra et~al.}(2008)\citenamefont{Mishra, Ram, Tiwari,
  and Mehendale}}]{mishra2}
\bibinfo{author}{\bibfnamefont{S.~R.} \bibnamefont{Mishra}},
  \bibinfo{author}{\bibfnamefont{S.~P.} \bibnamefont{Ram}},
  \bibinfo{author}{\bibfnamefont{S.~K.} \bibnamefont{Tiwari}},
  \bibnamefont{and} \bibinfo{author}{\bibfnamefont{S.~C.}
  \bibnamefont{Mehendale}}, \bibinfo{journal}{Phys. Rev. A}
  \textbf{\bibinfo{volume}{77}}, \bibinfo{pages}{065402}
  (\bibinfo{year}{2008}).

\bibitem[{\citenamefont{Muniz et~al.}(2001)\citenamefont{Muniz, Magalh\~aes,
  Courteille, Perez, Marcassa, and Bagnato}}]{muniz}
\bibinfo{author}{\bibfnamefont{S.~R.} \bibnamefont{Muniz}},
  \bibinfo{author}{\bibfnamefont{K.~M.~F.} \bibnamefont{Magalh\~aes}},
  \bibinfo{author}{\bibfnamefont{P.~W.} \bibnamefont{Courteille}},
  \bibinfo{author}{\bibfnamefont{M.~A.} \bibnamefont{Perez}},
  \bibinfo{author}{\bibfnamefont{L.~G.} \bibnamefont{Marcassa}},
  \bibnamefont{and} \bibinfo{author}{\bibfnamefont{V.~S.}
  \bibnamefont{Bagnato}}, \bibinfo{journal}{Phys. Rev. A}
  \textbf{\bibinfo{volume}{65}}, \bibinfo{pages}{015402}
  (\bibinfo{year}{2001}).

\bibitem[{\citenamefont{Cooper et~al.}(1994)\citenamefont{Cooper, Hillenbrand,
  Rink, Townsend, Zetie, and Foot}}]{cooper}
\bibinfo{author}{\bibfnamefont{C.~J.} \bibnamefont{Cooper}},
  \bibinfo{author}{\bibfnamefont{G.}~\bibnamefont{Hillenbrand}},
  \bibinfo{author}{\bibfnamefont{J.}~\bibnamefont{Rink}},
  \bibinfo{author}{\bibfnamefont{C.~G.} \bibnamefont{Townsend}},
  \bibinfo{author}{\bibfnamefont{K.}~\bibnamefont{Zetie}}, \bibnamefont{and}
  \bibinfo{author}{\bibfnamefont{C.~J.} \bibnamefont{Foot}},
  \bibinfo{journal}{Europhys. Lett.} \textbf{\bibinfo{volume}{28}},
  \bibinfo{pages}{397} (\bibinfo{year}{1994}).

\end{thebibliography}

\end{document}